\documentclass[aps,prd,amsmath,amssymb,reprint,
nofootinbib,
]{revtex4-1}

\usepackage{hepunits}
\usepackage{graphicx}\usepackage{dcolumn}\usepackage{bm}\usepackage[mathlines]{lineno}\usepackage{lipsum}
\usepackage{xspace}
\usepackage[nolist]{acronym}
\usepackage{color}
\usepackage[caption=false]{subfig}
\usepackage{enumitem}
\usepackage{comment}
\usepackage{subfig}
\usepackage{bm}
\usepackage[normalem]{ulem}

\graphicspath{{images/}}

\usepackage{etoolbox}
\makeatletter
\patchcmd\linenumberpar{\@LN@parpgbrk}{\penalty\@LN@parpgpen\relax}{}{}
\makeatother

\begin{document}
\title{Solutions to Axion Electrodynamics in Various Geometries}
\author{Jonathan Ouellet}
\email{ouelletj@mit.edu}
\affiliation{Laboratory for Nuclear Science, Massachusetts Institute of Technology, Cambridge, MA 02139, U.S.A.}

\author{Zachary Bogorad}
\affiliation{Laboratory for Nuclear Science, Massachusetts Institute of Technology, Cambridge, MA 02139, U.S.A.}

 \date{March 4, 2019}

\begin{abstract}
Recently there has been a surge of new experimental proposals to search for ultra-light axion dark matter with axion mass, $m_a\lesssim1\,\mu$eV. Many of these proposals search for small oscillating magnetic fields induced in or around a large static magnetic field. Lately, there has been interest in alternate detection schemes which search for oscillating electric fields in a similar setup. In this paper, we explicitly solve Maxwell's equations in a simplified geometry and demonstrate that in this mass range, the axion induced electric fields are heavily suppressed by boundary conditions. Unfortunately, experimentally measuring axion induced electric fields is not feasible in this mass regime using the currently proposed setups with static primary fields. We show that at larger axion masses, induced electric fields are not suppressed, but boundary effects may still be relevant for an experiment's sensitivity. We then make a general argument about a generic detector configuration with a static magnetic field to show that the electric fields are always suppressed in the limit of large wavelength.
\end{abstract}

\maketitle
\begin{acronym}
\acro{ABRA}[ABRACADABRA]{{\bf A} {\bf B}roadband/{\bf R}esonant {\bf A}pproach to {\bf C}osmic {\bf A}xion {\bf D}etection using an {\bf A}mplifying {\bf B}-field {\bf R}ing {\bf A}pparatus}
\acro{PSD}{power spectral density}
\acro{ALP}{axion-like particle}
\acro{ADM}[aDM]{axion dark matter}
\acro{WIMP}{Weakly Interacting Massive Particle}
\acro{MQS}{magneto-quasistatic}
\acro{DM}{Dark Matter}
\acro{SM}{Standard Model}
\acro{LSW}{light shining through wall}
\acro{TE}{Transverse Electric}
\acro{TM}{Transverse Magnetic}
\end{acronym}

\newcommand{\ABRA}{ABRACADABRA\xspace}
\newcommand{\abra}{ABRACADABRA-10\,cm\xspace}
\newcommand{\ALP}{\ac{ALP}\xspace}
\newcommand{\ALPs}{\ac{ALP}s\xspace}
\newcommand{\WIMP}{\ac{WIMP}\xspace}
\newcommand{\MQS}{\ac{MQS}\xspace}
\newcommand{\ADM}{\ac{ADM}\xspace}
\newcommand{\rhoDM}{\ensuremath{\rho_{\rm DM}}\xspace}
\newcommand{\gagg}{\ensuremath{g_{a\gamma\gamma}}\xspace}
\newcommand{\DM}{\ac{DM}\xspace}
\newcommand{\SM}{\ac{SM}\xspace}
\newcommand{\LSW}{\ac{LSW}\xspace}
\newcommand{\TE}{\ac{TE}\xspace}
\newcommand{\TM}{\ac{TM}\xspace}

\newcommand{\dt}[1]{\ensuremath{\frac{\partial {#1}}{\partial t}}}
\newcommand{\ddt}[1]{\ensuremath{\frac{\partial^2 {#1}}{\partial t^2}}}
\newcommand{\dal}[1]{\ensuremath{\nabla^2{#1}-\ddt{#1}}}
\newcommand{\dalf}[1]{\ensuremath{\nabla^2{#1}+\omega_a^2{#1}}}

\section{Introduction}

Starting about $10^4$ years after the Big Bang and lasting $10^{10}$ years after, the gravitational evolution of the universe was driven mostly by \DM. But despite the wealth of compelling evidence for \DM, we have not yet understood it on a particle level or determined how it fits next to the \SM of particle physics. In fact, the field of possible explanations for \DM is so broad as to incorporate masses from $\sim10^{-22}$\,eV to $\sim100M_{\odot}$.

One of the leading candidates to explain the \DM abundance is the axion. It was originally proposed to solve the strong-CP problem in QCD \cite{Peccei1977,Weinberg1977,Wilczek1977}, but its weak interaction strength with \SM particles and an elegant production mechanism in the early universe make it a promising candidate to explain \DM as well \cite{Preskill1983,Dine1983,Abbott1982}. 

Unlike the more thoroughly constrained \DM candidate, the \WIMP, the axion is expected to be extremely light with mass \mbox{$10^{-14}\lesssim m_a\lesssim1$\,eV} (see \cite{Essig2013l,Marsh2015,Graham2015} for a recent review). This implies that unlike \WIMP \DM, which would have a few particles per cubic meter, \ADM would have a very high number density and behave like a coherent field. In this case, the \DM energy density is better thought of as the kinetic and potential energy of a classical field rather than a dilute gas of individual particles.

If produced by the misalignment mechanism \cite{Preskill1983,Dine1983}, the time evolution of the axion field is expected to be given by
\begin{equation}
a(\mathbf{x},t) = a_0\cos(\omega_a t-\mathbf{x}\cdot\mathbf{k}_D)\,,
\end{equation}
where the frequency of oscillation is approximately equal to the axion mass $\omega_a=m_a$ and has an arbitrary overall phase. If \ADM is responsible for the observed \DM density, we can relate $a_0=\sqrt{2\rhoDM}/m_a$, where $\rhoDM$ is the local \DM density of $\sim0.3$\,GeV/cm$^3$ \cite{Read2014}. Though \ADM is extremely cold, it is expected to have a very small velocity spread due to gravitational effects. In the potential well of the Milky Way we expect a typical local velocity spread of $v_{\rm DM}\sim220$\,km/s. This results in a small spread in oscillation frequency due to Doppler shifting, \mbox{$\Delta\omega_a/\omega_a\sim v_{\rm DM}^2\approx10^{-6}$}, as well as small spatial gradients on the scale of the de~Broglie wavelength, $\lambda_D=2\pi/|\mathbf{k}_D|$.

Experiments searching for \ADM often leverage the fact that the axion couples to the photon and thus creates a small modification to electromagnetism. The axion -- or any \ALP\ for that matter -- will create a modification to the electromagnetic Lagrangian, that can be written in terms of the Maxwell field tensor $F^{\mu\nu}$, electric current $J_e^\mu$, and axion field $a$:
\begin{equation}
  \mathcal{L}_{\rm EM} = J^e_\mu A^{\mu} -\frac14 F^{\mu\nu}F_{\mu\nu}-\frac14\gagg a F^{\mu\nu}\tilde{F}_{\mu\nu}\,. \label{eqn:Lagrangian}
\end{equation}
Where $\tilde{F}_{\mu\nu}=\varepsilon_{\mu\nu\sigma\rho}F^{\sigma\rho}$, and \gagg is an unknown, but very small, coupling between the axion and photon.

The $aF\tilde{F}$ term can be treated as an axion-to-two-photon coupling which converts photons into axions and vice-versa, as in \LSW \cite{Spector:2016vwo} and axion helioscope \cite{CAST2017,Armengaud:2014gea} experiments. However, since \ADM would imply a high occupation number for the field $a$, an alternate approach is to write the ``axion current'',
\begin{equation}
J^\mu_a=\gagg\left(\mathbf{B}\cdot\bm{\nabla}a,-\mathbf{E}\times\bm{\nabla}a+\partial_ta \mathbf{B}\right)\,,
\label{eqn:AxionCurrent}
\end{equation}
which can then be easily incorporated into a modified form of Maxwell's equations \cite{Sikivie1983}
\begin{subequations}
\begin{eqnarray}
\bm{\nabla}\cdot \mathbf{E} &=& \rho-\gagg \mathbf{B}\cdot \bm{\nabla} a\,, \label{eqn:modmax1}\\
\bm{\nabla}\cdot \mathbf{B} &=& 0\,, \label{eqn:modmax2}\\
\bm{\nabla}\times \mathbf{E} &=& - \frac{\partial\mathbf{B}}{\partial t}\,, \label{eqn:modmax3}\\
\bm{\nabla}\times \mathbf{B} &=& \frac{\partial \mathbf{E}}{\partial t}+\mathbf{J}-\gagg\left(\mathbf{E}\times\bm{\nabla}a-\frac{\partial a}{\partial t}\mathbf{B}\right)\,. \label{eqn:modmax4}
\end{eqnarray}
\label{eqn:modmax}
\end{subequations}
A fifth equation  describes the evolution of the axion field, however, we will neglect it throughout this work as it only introduces corrections of order $\gagg^2$ or higher.\footnote{This is valid whenever $\frac{\gagg\mathbf{E}\cdot\mathbf{B}}{m_a\sqrt{\rho_{\rm DM}}}\ll1$, which is the case for the majority of axion haloscopes proposals. Though it is interesting to consider the case where it is not.}

A common type of axion haloscope experiment works by creating a strong static $\mathbf{B}$ field and looking for small AC fields driven at the frequency of the axion, $\omega_a$. As we will see below, the exact implications of these additional terms for an experiment will depend strongly on the relative size of the detector to oscillation wavelength $\lambda_a=2\pi/\omega_a$. Because for \ADM, the oscillation wavelength is almost exactly equal to the Compton wavelength, we will find that detectors searching for \ADM in different axion mass ranges will need to search for different electromagnetic effects corresponding to different oscillation wavelength limits. 

Experiments like ADMX \cite{Asztalos2001,Asztalos2009,ADMX2018}, HAYSTAC \cite{HAYSTAC2018a}, and others \cite{PhysRevD.42.1297,PhysRevLett.59.839,PhysRevLett.80.2043} utilize resonant cavities to probe axion masses in the range $m_a\sim10^{-6} - 10^{-5}$\,eV. In this range, the axion has Compton wavelength and therefore $\lambda_a$ of order $6-60$\,cm, comparable to the physical size of the detector. Practical considerations limit the range of masses that can be probed with detectors comparable in size to $\lambda_a$. At shorter wavelengths, $\lambda_a\sim1$\,mm, experiments like MADMAX propose to manipulate electric fields using arrays of dielectric plates \cite{TheMADMAXWorkingGroup:2016hpc} to coherently add effects over many Compton wavelengths within their detector. Recently, several experiments have been proposed to search for \ADM with with much lower masses of $10^{-14} - 10^{-6}$\,eV and therefore Compton wavelengths much larger than the detector. These include experiments like \ABRA \cite{ABRA2016}, DM Radio \cite{DMRadio_Design}, BEAST \cite{BEAST_Paper} and others \cite{Sikivie:2013laa,DeRocco:2018jwe,Obata:2018vvr,Liu:2018icu}.

In the limit of large $\lambda_a$, the typical approach is to build a detector with a strong DC magnetic field and search for an induced AC $\mathbf{B}$ field. Experiments like \cite{ABRA2016,DMRadio_Design,Sikivie:2013laa}, work in the \MQS regime -- equivalent to assuming that the displacement current in Eqn.~\ref{eqn:modmax4} is small. The axion term can then be treated as an effective current $\mathbf{J}_{\rm eff}$ that sources a real $\mathbf{B}$ field, which can be detected. However, \cite{BEAST_Paper} proposes an alternate approach, utilizing the displacement currents to measure an induced AC field $\mathbf{E}=-\gagg a \mathbf{B}$ in a strong DC $\mathbf{B}$ field. This has caused disagreement in the community about whether the axion induced electric field would be large enough to be observable or whether it is significantly suppressed -- specifically, whether the electric field is given by $\mathbf{E}=-\gagg a \mathbf{B}$, or whether it is suppressed by powers of $1/\lambda_a$. This has prompted new interpretations of the effect of the axion field in the presence of electromagnetic fields \cite{Tobar2018}. This debate has been further clouded by an old paper that directly calculates the induced Lorentz force on a test charge in the presence \ADM \cite{HONG1991} and appears to support the results in \cite{Tobar2018}. However, the calculation in that paper implicitly assumes a homogeneous $\mathbf{B}$ field, and neglects momentum transfer from virtual photons in the magnetic field and so is not so easily connected to a realistic experimental setup.

In Section~\ref{sec:TaylorExpansion}, we outline the field expansion approach we will use throughout this paper, and write down the modified wave equations in the presence of an axion field. In Section~\ref{sec:InfiniteSolenoid}, we explicitly solve the modified Maxwell's equations in the case of an infinite solenoid without assuming the \MQS approximation and demonstrate explicitly that the electric field is suppressed in the large $\lambda_a$ limit. In Section~\ref{sec:ArbitratyDistribution}, we generalize this conclusion and show that for a broad class of detectors, the \MQS approximation is always valid in the large $\lambda_a$ limit and that the suppression of the electric field is a generic quality. From this we conclude that in this mass regime, an experiment with a static $\mathbf{B}$ field will always be more sensitive to axion induced magnetic fields over electric fields. It is worth mentioning that this argument does not hold for experiments with time-varying primary fields, such as in recently proposed detection schemes based on interferometry \cite{Obata:2018vvr,DeRocco:2018jwe,Liu:2018icu}. Finally, in Section~\ref{sec:VacuumPolarization}, we discuss the alternate -- but completely equivalent -- approach outlined in \cite{Tobar2018}, and the physical intuition it can provide.

As is common, we will assume that the spatial gradients of the axion field are negligible, $\bm{\nabla} a\approx0$. This is because the de~Broglie wavelength is about three orders of magnitude larger than the oscillation wavelength ($\lambda_D\approx10^3\lambda_a$), and thus spatial gradient terms are suppressed.

\section{Field Expansions}
\label{sec:TaylorExpansion}

Our general approach throughout this paper will be to first Taylor expand $\mathbf{E}$ and $\mathbf{B}$ in powers of \gagg and then convert Eqns.~\ref{eqn:modmax} into wave equations which can then be grouped into like powers of \gagg. 

We consider a detector with primary fields driven by a \emph{static} current $\mathbf{J}_e(\mathbf{x})$ and charge distribution $\rho_e(\mathbf{x})$. For example, these can be thought as currents driving a magnet or producing an electric field and driven externally by a power supply. For simplicity, we calculate only the behavior of the vacuum $\mathbf{E}$ and $\mathbf{B}$ fields in response to $\mathbf{J}_e(\mathbf{x})$, $\rho_e(\mathbf{x})$ and the axion field $a$. However, it is straightforward to include the response of free or bound charges through the usual use of additional $\rho,\mathbf{J}$ terms or using the macroscopic $\mathbf{D}$ and $\mathbf{H}$ fields.

Because $\gagg$ is expected to be very small, we can expand\footnote{In fact, we utilize the fact that $J_a^\mu\ll J_e^\mu$ to Taylor expand the effects of the axion induced $\mathbf{E}$ and $\mathbf{B}$ about the primary fields. But we can use \gagg to keep track of the order of the expansion.} the $\mathbf{E}$ and $\mathbf{B}$ fields into terms of equal order in \gagg:
\begin{subequations}
\begin{eqnarray}
\mathbf{E}(\mathbf{x},t)=\mathbf{E}_0(\mathbf{x},t)+\mathbf{E}_1(\mathbf{x},t)+O(\gagg^2)\,,\label{eqn:TaylorExpansionE}\\
\mathbf{B}(\mathbf{x},t)=\mathbf{B}_0(\mathbf{x},t)+\mathbf{B}_1(\mathbf{x},t)+O(\gagg^2)\,,\label{eqn:TaylorExpansionB}
\end{eqnarray}
\label{eqn:TaylorExpansion}
\end{subequations}
where $\mathbf{E}_1$ and $\mathbf{B}_1$ will be proportional to \gagg.

We can take the time derivative of Eqn.~\ref{eqn:modmax4}, and group equations in constant powers of \gagg to get
\begin{subequations}
\begin{eqnarray}
\bm{\nabla}^2\mathbf{E}_0(\mathbf{x},t)&=&\ddt{\mathbf{E}_0(\mathbf{x},t)}+\bm{\nabla}\rho_e(\mathbf{x})\,,\label{eqn:InfSolWaveEquationsE0}\\
\bm{\nabla}^2\mathbf{E}_1(\mathbf{x},t)&=&\ddt{\mathbf{E}_1(\mathbf{x},t)}+\gagg\ddt{a}\mathbf{B}_0(\mathbf{x},t)\,,\label{eqn:InfSolWaveEquationsE1}
\end{eqnarray}
\end{subequations}
where we have taken advantage of the fact that 
\begin{eqnarray}
-\bm{\nabla}\times(\bm{\nabla}\times\mathbf{E}) &=& \bm{\nabla}^2\mathbf{E}-\bm{\nabla}(\bm{\nabla}\cdot\mathbf{E}) \nonumber\\
&=&\bm{\nabla}^2\mathbf{E}-\bm{\nabla}\rho_e\,,
\end{eqnarray}
and assumed that $\rho_e$ and $\mathbf{J}_e$ are constant in time. We have also dropped the terms of order $\gagg^2$ or higher.

Similarly, we could take the time derivative of Eqn.~\ref{eqn:modmax3} and group in like powers of \gagg, to get
\begin{subequations}
\begin{equation}
-\bm{\nabla}^2\mathbf{B}_0(\mathbf{x},t)+\bm{\nabla}\times\mathbf{J}_e(\mathbf{x})=-\ddt{\mathbf{B}_0}(\mathbf{x},t)
\end{equation}
\begin{equation}
-\bm{\nabla}^2\mathbf{B}_1(\mathbf{x},t)+\gagg\dt{a}\bm{\nabla}\times\mathbf{B}_0(\mathbf{x},t) =-\ddt{\mathbf{B}_1(\mathbf{x},t)}\,.
\end{equation}
\end{subequations}

Combining these equations, we are left with the wave equations to solve:
\begin{subequations}
\begin{eqnarray}
\dal{\mathbf{E}_0(\mathbf{x},t)} &=& \bm{\nabla} \rho_e(\mathbf{x}) \label{eqn:WaveEquationE0}\\
\dal{\mathbf{B}_0(\mathbf{x},t)} &=& -\bm{\nabla}\times\mathbf{J}_e(\mathbf{x})\label{eqn:WaveEquationB0}\\
\dal{\mathbf{E}_1(\mathbf{x},t)} &=& \gagg\ddt{a}\mathbf{B}_0(\mathbf{x},t)
\label{eqn:WaveEquationE1}\\
\dal{\mathbf{B}_1(\mathbf{x},t)} &=&\gagg\dt{a}\bm{\nabla}\times\mathbf{B}_0(\mathbf{x},t). \label{eqn:WaveEquationB1}
\end{eqnarray}
\label{eqn:WaveEquations}
\end{subequations}
To reiterate, at this point we have assumed only that the primary fields $\mathbf{E}_0$ and $\mathbf{B}_0$ are static, i.e. that $\rho_e$ and $\mathbf{J}_e$ are constant in time. We have also neglected the gradient of the axion field. Below, we use this expansion to examine the effect of the axion field on particular choices of geometry. 
\section{Axion Dark Matter and the Infinite Solenoid}
\label{sec:InfiniteSolenoid}

\begin{figure}
\centering
\includegraphics[width=.3\textwidth]{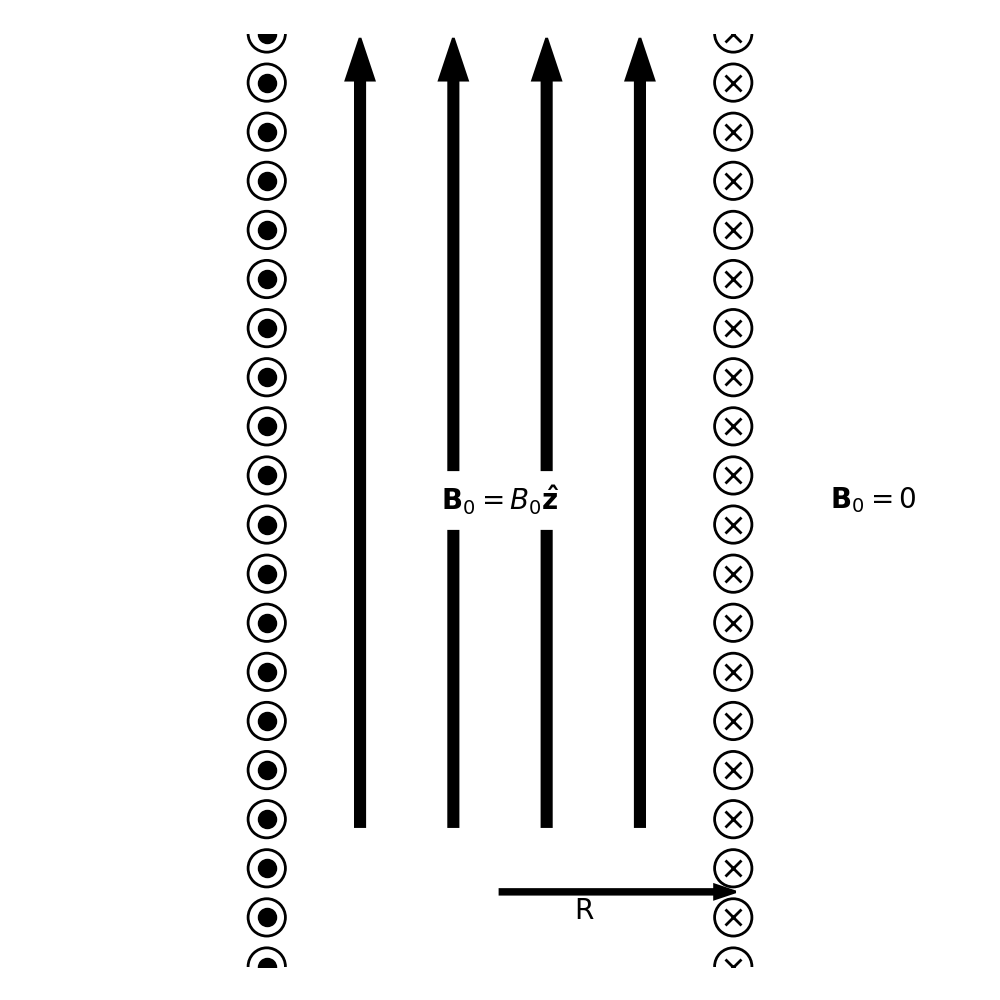}
\caption{Diagram of a simplified geometry with an infinite solenoid pointing along the $\mathbf{\hat{z}}$ direction. The solution without an axion is identically $B_0\mathbf{\hat{z}}$ inside and 0 outside.}
\label{fig:infinite_solenoid}
\end{figure}

The simplest geometry to consider is the case of the infinitely tall solenoid. Of course, in practice this geometry is not physically achievable. A physical solenoid will have a finite extent and thus returning fields outside the winds of the solenoid. But in many experimental setups, these fringe fields are small compared to the field inside the solenoid and lead to sub-dominant corrections. An infinite solenoid is a useful example on which to see the major effects. 

Assume we have an infinitely tall solenoid of radius $R$ pointing
along the $\mathbf{\hat{z}}$ direction. In cylindrical coordinates $(\rho,\phi,z)$, the current density
along the walls is
\begin{equation}
\mathbf{J}_e = B_0\delta(\rho-R)\bm{\hat{\phi}}\,,
\end{equation}
such that the unmodified Maxwell's equations would lead to the solution
\begin{equation}
\mathbf{B}_0 = 
\begin{cases}
B_0\mathbf{\hat{z}} & \rho<R\,, \\
0 & \rho>R\,.
\end{cases}
\label{eqn:InfSolAxionFreeSolution}
\end{equation}
See Fig.~\ref{fig:infinite_solenoid}. Further, let's assume that current cannot flow along the solenoid walls in the $\mathbf{\hat{z}}$ direction. For instance, we can take this to be a densely packed set of current carrying loops that only carry current in the $\mathbf{\hat{\phi}}$ direction. Further we assume $\rho_e=0$, such that $\mathbf{E}_0=0$.

In this geometry, Eqns.~\ref{eqn:WaveEquationE0} and \ref{eqn:WaveEquationB0} reproduce the classical result to zeroth order in \gagg, given in Eqn.~\ref{eqn:InfSolAxionFreeSolution}.
We can then rewrite Eqns.~\ref{eqn:WaveEquationE1} and \ref{eqn:WaveEquationB1} as
\begin{subequations}
\begin{eqnarray}
\dal{\mathbf{E}_1} &=& \begin{cases} \gagg\ddt{a}B_0\mathbf{\hat{z}} & \rho < R\,,\\
0 & \rho>R\,, \end{cases}\label{eqn:WaveEquationInfSolE1}\\
\dal{\mathbf{B}_1} &=& -\gagg\dt{a}B_0\delta(\rho-R)\bm{\hat{\phi}}\,.\label{eqn:WaveEquationInfSolB1}
\end{eqnarray}
\end{subequations}
It is clear from these equations that the only non-trivial solutions will be for $E_z$ and $B_\phi$. The other components are not affected by the axion field at leading order. Since the axion field is nicely decomposable into frequency modes, we will move into frequency space and drop transient solutions. Because of the symmetry, we propose the solutions
\begin{subequations}
\begin{eqnarray}
E_{1z}(\rho,t) &=& \psi_E(\rho)e^{i\omega_a t}\,, \label{eqn:prop_sol_E}\\
B_{1\phi}(\rho,t) &=& \psi_B(\rho)e^{i\omega_a t} \,. \label{eqn:prop_sol_B}
\end{eqnarray}
\end{subequations}

\subsection{The \bf{B} field Solution}

Plugging (\ref{eqn:prop_sol_B}) into (\ref{eqn:WaveEquationInfSolB1}) and performing a change of variables to $\rho'=\omega_a\rho$, we get the Bessel equation with a boundary condition at $\rho=R$:
\begin{eqnarray}
\left(\partial^2_{\rho'}+\frac{1}{\rho'}\partial_{\rho'}+\left(1-\frac{1}{\rho'^2}\right)\right)\psi_B =\\ -i\gagg a_0 B_{0}\delta(\rho'-\omega_aR)\,.&\nonumber
\end{eqnarray}
The solutions to this are Bessel functions of order 1,
with boundary conditions at $\rho'=0$ and $\rho'=\omega_aR$:
\begin{equation}
\psi_B(\rho')=\begin{cases}a_BJ_1(\rho') & \rho' < \omega_aR\,,\\ b_B H^+_1(\rho') & \rho'>\omega_aR\,. \end{cases} \label{eqn:Bessel_Sol_B}
\end{equation}
Here, we required that for $\rho'<\omega_aR$ the diverging
$N_1(\rho')$ solution is suppressed, and for $\rho'>\omega_aR$ an
outward traveling wave given by the Hankel function, $H^+_1(\rho')$. (An inward traveling wave, $H^-_1(\rho')$, is also a correct solution, however would imply power flowing into the oscillating axion field from infinity rather than out of it.)

We can now find the full solution, by requiring continuity of $B_{\parallel}$ across the boundary, and a step discontinuity in $\frac{\partial \psi_B}{\partial \rho'}$ as required by the $\delta$ function. (Remember that we specified that current could not flow along $\mathbf{\hat{z}}$.)
\begin{subequations}
\begin{eqnarray}
a_BJ_1(\omega_aR) - b_BH^+_1(\omega_aR) =0 \label{eqn:bound_cond_B_1}\,, \\
\left.\left(b_B\partial_{\rho'}H^+_1(\rho')-a_B\partial_{\rho'}J_a(\rho')\right)\right|_{\rho'=\omega_aR} =\\-i\gagg a_0B_0\,.\nonumber
\end{eqnarray}
\end{subequations}
This can then be solved further to yield
\begin{subequations}
\begin{eqnarray}
a_B = -\frac\pi2\gagg a_0B_0\omega_aRH_1^+(\omega_aR)\,,\\
b_B = -\frac\pi2\gagg a_0B_0\omega_aRJ_1(\omega_aR)\,,
\end{eqnarray}
\end{subequations}
where we have leveraged Abel's identity to simplify the Wronksian of Bessel functions as
\begin{eqnarray}
\mathcal{W}(J_1,H_1^+)=J_1\frac{\partial H_1^+}{\partial \rho'}-\frac{\partial J_1}{\partial \rho'}H_1^+=\frac{2i}{\pi\rho'}\,.
\end{eqnarray}
This fully specifies the solution of the $\mathbf{B}$ field driven by the axion at leading order in \gagg. Figure~\ref{fig:infinite_sol_field_strength} shows the behavior of $B_{1\phi}$ for various values of $R/\lambda_a$. 

\subsection{The \bf{E} field Solution}

Returning to the $E_{1z}$ component, we can plug (\ref{eqn:prop_sol_E}) into (\ref{eqn:WaveEquationInfSolE1}) and performing a change of variables get another Bessel equation:
\begin{eqnarray}
\left(\partial^2_{\rho'}+\frac{1}{\rho'}\partial_{\rho'} +1\right)\psi_E&=&\begin{cases}-\gagg a_0B_0 & \rho' < \omega_aR\,,\\0 &\rho'>\omega_aR\,,\end{cases} \label{eqn:Solenoid_Bessel_Eqn_E}
\end{eqnarray}
which has solutions
\begin{equation}
\psi_E(\rho')=\begin{cases} a_E J_0(\rho') -\gagg a_0B_0& \rho' < \omega_aR\,, \\
b_EH^+_0(\rho') & \rho'>\omega_aR\,. \end{cases} \label{eqn:Bessel_Sol_E}
\end{equation}
Again, we have required that $\psi_E(\rho')$ be finite at $\rho'=0$, and an outward traveling wave for $\rho'>\omega_aR$.

Here, the boundary conditions require that $E_z$ and its derivative be continuous across the boundary. The former condition can be seen by integrating $\bm{\nabla}\times\mathbf{E}$ around a small contour just inside and outside of the solenoid; the latter can be seen by integrating Eqn.~\ref{eqn:Solenoid_Bessel_Eqn_E} between $[\omega_aR-\varepsilon,\omega_aR+\varepsilon]$ as $\varepsilon\rightarrow0$.
\begin{subequations}
\begin{eqnarray}
a_EJ_0(\omega_aR)-\gagg a_0B_0 &=& b_EH^+_0(\omega_aR)\,, \label{eqn:bound_cond_E_1}\\
a_EJ_1(\omega_aR)&=&b_EH_1^+(\omega_aR)\,.\label{eqn:bound_cond_E_2}
\end{eqnarray}
\label{eqn:bound_cond_E}
\end{subequations}
We can again simplify this further to
\begin{subequations}
\begin{eqnarray}
a_E=\frac{i\pi}{2}\gagg a_0B_0\omega_aRH_1^+(\omega_aR)\,,\\
b_E=\frac{i\pi}{2}\gagg a_0B_0\omega_aRJ_1^+(\omega_aR)\,.
\end{eqnarray}
\end{subequations}
Where we have taken advantage of the Bessel function property that $2\partial_{\rho'}\Omega_\nu=\Omega_{\nu-1}-\Omega_{\nu+1}$ and that $\Omega_{-\nu}=(-1)^{\nu}\Omega_\nu$ for $\Omega_\nu\subset[J_\nu,H^+_{\nu}]$.
These equations fully specify the $\mathbf{E}$ field solution.

\begin{figure*}
\centering
\includegraphics[width=.8\textwidth]{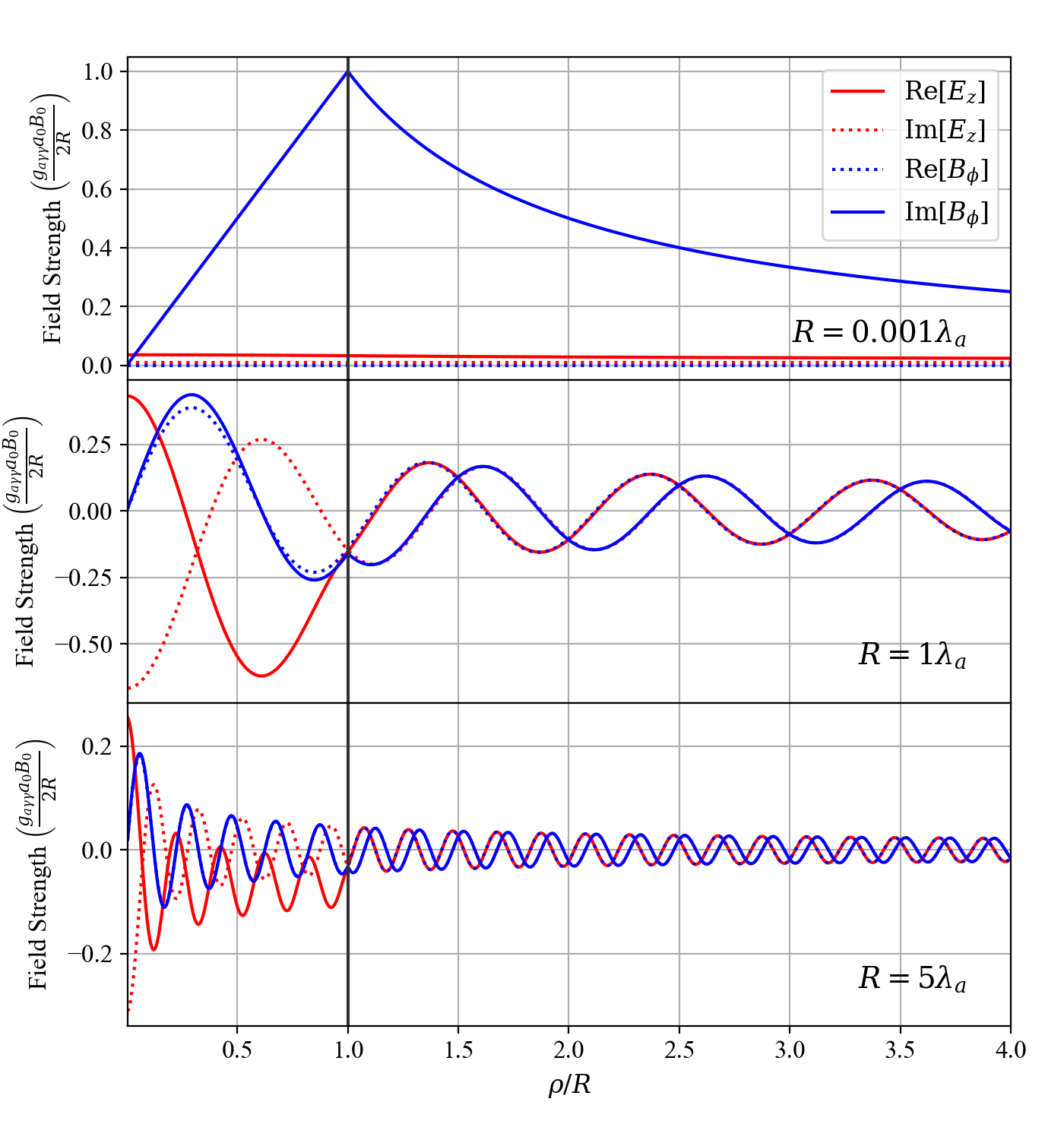}
\caption{Analytic solutions for the field strengths for the infinite solenoid configuration. The $E_{1z}$ and $B_{1\phi}$ field strengths are plotted in units of $\frac{\gagg a_0B_0}{2R}$ for several values of $R/\lambda_a$. The only approximation are that these are to first order in \gagg.}
\label{fig:infinite_sol_field_strength}
\end{figure*}

Putting these together with the solutions for the $\mathbf{B}$ field yields a nice compact form
\begin{eqnarray}
\begin{pmatrix}
a_E\\b_E\\a_B\\b_B
\end{pmatrix}
=\frac{\pi\gagg a_0B_0\omega_aR}{2}
\begin{pmatrix}
iH_1^+(\omega_aR)\\ iJ_1(\omega_aR)\\ -H_1^+(\omega_aR)\\-J_1(\omega_aR) 
\end{pmatrix}\,.
\label{eqn:Matrix_of_coefficients}
\end{eqnarray}
The solutions for $E_{1z}$ and $B_{1\phi}$ are plotted together in Fig.~\ref{fig:infinite_sol_field_strength} for various values of $R/\lambda_a$. It is worth pointing out that this is in fact the solution to an infinite wire with an ``effective current'' given by $\mathbf{J}_{\rm eff}=\gagg \dt{a}B_0\mathbf{\hat{z}}$.

\subsection{The Long Wavelength Limit}
\label{sec:InfiniteSolenoidLongWavelength}
The variable $\rho'$, is actually the ratio of the radial coordinate scaled by the oscillation wavelength of the axion $\rho' = 2\pi\rho/\lambda_a$. Not surprisingly, this marks this wavelength as the relevant length scale of the problem. If $R\ll\lambda_a$, we will get one type of behavior, as compared to $R\sim\lambda_a$ or $R\gg\lambda_a$. This can be seen in Fig.~\ref{fig:infinite_sol_field_strength}.

In the long $\lambda_a$ limit, $R\ll\lambda_a$ (or equivalently $\rho'=\omega_aR\ll1$), both sides the solenoid can be thought of as ``oscillating in phase'' and the fields add coherently over the relevant distance scales. This is the limit relevant for experiments like \ABRA \cite{ABRA2016}, DM Radio \cite{DMRadio_Design}, BEAST \cite{BEAST_Paper} and other LC-resonator searches \cite{Sikivie:2013laa}.

We can take the asymptotic limits of the Bessel functions to see how the field near the solenoid behaves. Equation~\ref{eqn:Bessel_Sol_B} becomes
\begin{equation}
\psi_B(\rho')\approx\begin{cases} \frac{a_B}{2}\rho' & \rho'<\omega_aR\,, \\
-i\frac{2b_B}{\pi\rho'} & \rho'>\omega_aR\,.
\end{cases}
\end{equation}
with the coefficients given by
\begin{subequations}
\begin{eqnarray}
a_B &=& i\gagg a_0B_0\,,\\
b_B &=& -\frac{\pi\gagg a_0B_0\omega_a^2R^2}{4}\,,
\end{eqnarray}
\end{subequations}
inserting this and converting back to $\rho$, yields the radial behavior 
\begin{equation}
\psi_B(\rho) \approx \begin{cases} \frac{i}{2}\gagg \omega_aa_0 B_0\rho & \rho< R\,,\\
\frac{i}{2}\gagg \omega_a a_0 B_0 \frac{R^2}{\rho} & \rho>R\,.\end{cases}
\end{equation}
The factor of $i$ simply indicates a $\frac\pi2$-phase shift from the axion field. This is expected since the $\mathbf{B}$ field in Eqn.~\ref{eqn:WaveEquationB1} is driven by $\dt{a}$.

Plugging this back into Eqn.~\ref{eqn:prop_sol_B}, we have our full solution for the axion induced $\mathbf{B}$ field to first order in \gagg and in the limit of $\rho,R\ll\lambda_a$:
\begin{equation}
\mathbf{B}_1(\mathbf{x},t) \approx \begin{cases} \frac12\gagg \dt{a}\rho\bm{\hat{\phi}} & \rho<R\,,\\
\frac12\gagg \dt{a}\frac{R^2}{\rho}\bm{\hat{\phi}} & \rho>R\,. \end{cases}
\end{equation}
Here, we have summed over axion frequency modes $\omega_a$ to convert $i\omega_a a_0e^{i\omega_at}$ back into $\dt{a}$ to make the solution true for arbitrary $a(t)$. 

It should be noted, that this is exactly the result that we would expect from taking the \MQS approximation as is done in \cite{ABRA2016,DMRadio_Design,Sikivie:2013laa}.

Looking at the electric field behavior in the long wavelength limit, Eqn.~\ref{eqn:Bessel_Sol_E} becomes 
\begin{equation}
\psi_E(\rho') \approx \begin{cases} a_E\left(1-\frac{\rho'^2}{4}\right)-\gagg a_0B_0 & \rho' < \omega_aR \\
b_E\frac{2i}{\pi}\gamma'(\rho')&\rho' > \omega_aR 
\end{cases}
\end{equation}
where we define the function $\gamma'(x)=\ln(x/2)+\gamma-i\pi/2$, where $\gamma$ is the Euler-Mascheroni constant, ($\gamma\approx0.5772...$). With the coefficients given by
\begin{eqnarray}
\small
a_E &=& \gagg a_0B_0\nonumber\\&&\hspace{2mm}-\frac{\gagg a_0B_0\omega_a^2R^2}{2}\left(\gamma'(\omega_aR)-\frac12\right)
\,,\label{eqn:Bessel_Sol_E_Limit_Inside}\\
b_E &=& \frac{i\pi\gagg a_0B_0\omega_a^2R^2}{4}\,.
\end{eqnarray}
Expanding this out, and dropping terms of order $(\omega_aR)^2(\omega_a\rho)^2$, we can write
\begin{equation}
\small
\psi_E(\rho)=-\frac12\gagg a_0B_0(\omega_aR)^2\begin{cases}
\left(\gamma'(\omega_aR)-\frac12\right)+\frac{\rho^2}{2R^2} & \rho < R\,,\\
\gamma'(\omega_a\rho) & \rho>R\,.
\end{cases}
\end{equation}
This implies that, to first order in $\gagg$ and for $\rho,R\ll\lambda_a$, electric fields are suppressed by $\left(\frac{R}{\lambda_a}\right)^2\ln\left(\frac{R}{\lambda_a}\right)\ll\frac{R}{\lambda_a}$.
This behavior can be seen in Fig.~\ref{fig:infinite_sol_field_strength}.

This is in direct contrast with the argument set forth in \cite{BEAST_Paper}, which searches for an axion induced electric field in the long oscillation wavelength limit inside the solenoid. This conclusion is reached here using a particular geometry, but the conclusion is a lot more general, as we will show in the next section. It is worth noting that the $\mathbf{E}$ field solution proposed in that work, $\mathbf{E}=-\gagg a\mathbf{B}_0$, does appear in the solution to Maxwell's equations as the $\rho'$ independent term in Eqn.~\ref{eqn:Bessel_Sol_E}. But in the large $\lambda_a$ limit it is canceled by the  other term in the full solution -- given in Eqn.~\ref{eqn:Bessel_Sol_E_Limit_Inside}. 

In the short oscillation wavelength limit, the field $\mathbf{E}=-\gagg a\mathbf{B}_0$ appears as an offset to the oscillating Bessel function: \mbox{$E_z=\left(a_EJ(\omega_a\rho)-\gagg a_0B_0\right)e^{i\omega_a t}$}. When the Bessel function has many oscillations within $0<\rho<R$, the spatial average approaches $-\gagg a_0B_0e^{i\omega_a t}$. This can be seen in the lower panel of Fig.~\ref{fig:infinite_sol_field_strength} as the offset between the solid and dotted red lines. 

An experimental setup with a capacitor inside the solenoid (similar to \cite{BEAST_Paper}) would in fact see charges displaced by the oscillating axion induced $\mathbf{E}$ field. But this would only be a measurable effect in the $R\gtrsim\lambda_a$ limit (i.e. for frequencies $\omega_a/(2\pi)\gtrsim300$\,MHz). This is akin to the microwave cavity designs used by \cite{Asztalos2001,Asztalos2009,ADMX2018,HAYSTAC2018a,PhysRevD.42.1297,PhysRevLett.59.839,PhysRevLett.80.2043}, but without the resonator cavity. Interestingly, there are other recent proposals for the $R\sim\lambda_a$ regime using this type of detector, but with all resonant enhancement moved into electronics \cite{Daw2018}. At shorter wavelengths still, other experimental techniques have been proposed which rely on manipulating the $\mathbf{E}$ field with dielectric plates.\cite{TheMADMAXWorkingGroup:2016hpc}. These latter approaches, where $R\gtrsim\lambda_a$, are not incompatible with the results presented here.

\section{Demonstrating the MQS Approximation for a Generic Detector}
\label{sec:ArbitratyDistribution}

The argument in the previous section can be made much more general by directly demonstrating that the \MQS approximation holds in the presence of an oscillating axion field in the large $\lambda_a$ limit. In the following argument, we will make two assumptions:
\begin{enumerate}
\item our detector is composed of a collection of \emph{time-independent} charges and currents, $\rho_e$ and $\mathbf{J}_e$;
\item our detector fits into some box with a diagonal size $L$. Thus both the $\rho_e$ and $\mathbf{J}_e$ used to create our primary fields and whatever apparatus we use to detect axion induced fields are contained within $|\mathbf{x}-\mathbf{x}'|<L$.
\end{enumerate}
The precise shape of the box in the second assumption is irrelevant -- it only establishes a characteristic size for our detector. We make no assumptions about the configuration of the currents and charges within the box. 

We first convert the right hand sides of  Eqns.~\ref{eqn:WaveEquations} to include only terms of $\rho_e$ and $\mathbf{J}_e$ instead of $\mathbf{E}_0$ and $\mathbf{B}_0$. This is because, while the latter two fields can extend beyond the box, the second assumption above contains the charges and currents inside the box and therefore that they are zero on the surface of the box.

At this point, it is clear that the primary fields $\mathbf{E}_0$ and $\mathbf{B}_0$ (which are solutions to the axion-free equations) will be independent of time. We can then Fourier decompose $\mathbf{E}_1$ and $\mathbf{B}_1$ in the frequency domain:
\begin{subequations}
\begin{eqnarray}
\mathbf{E}_1(\mathbf{x},t)&=&\mathbf{E}_1(\mathbf{x})e^{i\omega_a t}\,,\\
\mathbf{B}_1(\mathbf{x},t)&=&\mathbf{B}_1(\mathbf{x})e^{i\omega_a t}\,,
\end{eqnarray}
\end{subequations}
and write the following wave equations
\begin{subequations}
\begin{eqnarray}
\bm{\nabla}^2\mathbf{E}_0(\mathbf{x}) &=& \bm{\nabla}\rho_e(\mathbf{x})\,,\label{eqn:WaveFEquationE0}\\
\bm{\nabla}^2\mathbf{B}_0(\mathbf{x}) &=& -\bm{\nabla}\times\mathbf{J}_e(\mathbf{x})\,,\label{eqn:WaveFEquationB0}\\
\dalf{\mathbf{E}_1(\mathbf{x})} &=& -\gagg \omega_a^2a_0\mathbf{B}_0(\mathbf{x})\,,\label{eqn:WaveFEquationE1}\\
\dalf{\mathbf{B}_1(\mathbf{x})} &=& i\gagg \omega_a a_0\bm{\nabla}\times\mathbf{B}_0(\mathbf{x})\,.\label{eqn:WaveFEquationB1}
\end{eqnarray}
\end{subequations}
We can trivially rewrite the RHS of Eqn.~\ref{eqn:WaveFEquationB1} in terms of $\mathbf{J}_e$ using Eqn.~\ref{eqn:WaveFEquationB0}. Focusing on Eqn.~\ref{eqn:WaveFEquationE1}, we can split $\mathbf{E_1}(\mathbf{x})$ into $\mathbf{E}_1(\mathbf{x})=\mathbf{E}_1'(\mathbf{x})-\gagg a_0\mathbf{B}_0(\mathbf{x})$, and get an equation for $\mathbf{E}_1'$ 
\begin{eqnarray}
\dal{\mathbf{E}'_1(\mathbf{x})} &=&\gagg a_0\bm{\nabla}^2\mathbf{B}_0(\mathbf{x})\nonumber\\
&=&\gagg a_0\bm{\nabla}\times\mathbf{J}_e(\mathbf{x})\,.
\end{eqnarray}

At this point, we can use the retarded Green's functions to solve for our fields. 
\begin{subequations}
\begin{eqnarray}
\mathbf{E}_0(\mathbf{x})&=&\frac{1}{4\pi}\int\frac{\bm{\nabla}\rho_e}{|\mathbf{x}-\mathbf{x}'|}\,d^3\mathbf{x}'\,,\label{eqn:greens_func_sol_E0}\\
\mathbf{B}_0(\mathbf{x})&=&\frac{1}{4\pi}\int\frac{\bm{\nabla}\times\mathbf{J}_e(\mathbf{x'})}{|\mathbf{x}-\mathbf{x}'|}\,d^3\mathbf{x}'\,,\label{eqn:greens_func_sol_B0}\\
\mathbf{E}_1(\mathbf{x})&=&\frac{\gagg a_0}{4\pi}\int\frac{e^{i\omega_a|\mathbf{x}-\mathbf{x}'|}-1}{|\mathbf{x}-\mathbf{x}'|}\bm{\nabla}\times\mathbf{J}_e(\mathbf{x}')\,d^3\mathbf{x}'\,,\label{eqn:greens_func_sol_E1}\\
\mathbf{B}_1(\mathbf{x})&=&\frac{i\gagg \omega_aa_0}{4\pi}\int\frac{e^{i\omega_a|\mathbf{x}-\mathbf{x}'|}}{|\mathbf{x}-\mathbf{x}'|}\mathbf{J}_e(\mathbf{x}')\,d^3\mathbf{x}'\,.\label{eqn:greens_func_sol_B1}
\end{eqnarray}
\end{subequations}
Notice that the $-1$ in the Eqn.~\ref{eqn:greens_func_sol_E1} came from solving for $\mathbf{E}_1'$ and substituting Eqn.~\ref{eqn:greens_func_sol_B0} in for the offset term, $-\gagg a_0 \mathbf{B}_0$.

We point out that $\rho_e$ does not appear in our axion induced fields. This is because in the limit that $\bm{\nabla} a$ is small, we cannot use static electric fields alone to detect axions -- regardless of their shape. This is evident from Eqns.~\ref{eqn:modmax}.

At this point our solution is very general. It is worth noticing the similarity between the solutions for $\mathbf{E}_1$ and $\mathbf{B}_1$ and the solutions to a multipole antenna. Equations~\ref{eqn:greens_func_sol_E1} and \ref{eqn:greens_func_sol_B1} are exactly the solutions to a current excitation of the form $\mathbf{J}_a=\gagg \partial_t a \mathbf{B}_0$, thus justifying the treatment of the axion induced effects as a effective, current to leading order in \gagg.

Up to now, we have  only used the first assumption that our charges and currents are constant in time. We use the second assumption to examine what happens in the limit of $L\ll\lambda_a$. Notice that our solutions are completely in terms of charges and currents, which are completely contained within our box of size $L$ -- as opposed to fields, which can extend outside of the box.

If both $\mathbf{x}$ and $\mathbf{x}'$ are within our box then $|\mathbf{x}-\mathbf{x}'|\le L$. And now we examine the behavior of the axion induced electric fields by Taylor expanding Eqn.~\ref{eqn:greens_func_sol_E1} in the limit of $\omega_aL\ll1$, and keeping first order terms:
\begin{eqnarray}
\mathbf{E}_1(\mathbf{x})&\approx&\frac{\gagg a_0}{4\pi}\int\frac{i\omega_a|\mathbf{x}-\mathbf{x}'|+O((\omega_a L)^2)}{|\mathbf{x}-\mathbf{x}'|}\bm{\nabla}\times\mathbf{J}_e(\mathbf{x}')\,d^3\mathbf{x}'\nonumber\\
&=&\frac{\gagg a_0 \omega_a}{4\pi} \int \bm{\nabla}\times\mathbf{J_e}(\mathbf{x}')\,d^3\mathbf{x}'+O((\omega_a L)^2)\nonumber\\
&=&\frac{\gagg a_0\omega_a}{4\pi}\int_S\mathbf{\hat{n}}\times\mathbf{J}_e(\mathbf{x}')dA'+O((\omega_a L)^2)\nonumber\\
&=&O((\omega_a L)^2)\hspace{7mm}(L\ll\lambda_a)\,.
\label{eqn:GenericLongWavelengthLimitE}
\end{eqnarray}
Where the surface integral vanishes due the fact that our current is contained within $S$ and so is equal to zero at the surface. Hence, the electric field is suppressed by $(L/\lambda_a)^2$.

Of course, a similar process can be done for Eqn.~\ref{eqn:greens_func_sol_B1}, but it is easy to see that the relevant difference between this equation and Eqn.~\ref{eqn:greens_func_sol_E1} is the $-1$ in the numerator. The leading term remains and the result is not suppressed by an additional powers of $\lambda_a$.

This conclusion is very general and does not depend on the precise details of our detector. We only assumed that 1) the currents and charges that drive our primary fields are constant in time; and 2) our detector is of characteristic size $L\ll \lambda_a$. Under these assumptions we have shown that axion induced electric fields are \emph{always} suppressed. We have actually just showed that the \MQS approximation continues to hold in the presence of an oscillating axion field with large $\lambda_a$. 

An interesting thing worth noting is that in this calculation we have neglected terms proportional to $\bm{\nabla}a$ as they are suppressed factors of $\lambda_a/\lambda_D\sim10^{-3}$. However, when $L/\lambda_a\lesssim10^{-3}$, it is possible for electric fields generated by the $\bm{\nabla}a\cdot\mathbf{B}$ term in Eqn.~\ref{eqn:modmax1} to dominate over the electric fields generated by the $\dt{a}\mathbf{B}$ term in Eqn.~\ref{eqn:modmax4}. 

Finally, it is worth describing the behavior of $\mathbf{E}_1$ and $\mathbf{B}_1$ in the limit that $\lambda_a$ is small compared to all other length scales. In this limit, the exponentials in Eqns.~\ref{eqn:greens_func_sol_E1} and~\ref{eqn:greens_func_sol_B1} oscillate very rapidly and will cause the integrals to average to zero. All that will remain is 
\begin{equation}
\mathbf{E}_1(\mathbf{x})\approx-\frac{\gagg a_0}{4\pi}\int\frac{\bm{\nabla}\times\mathbf{J}_e(\mathbf{x'})}{|\mathbf{x}-\mathbf{x}'|}\,d^3\mathbf{x}' \hspace{7mm} (L\gg\lambda)\,,
\end{equation}
which is exactly the $-\gagg a \mathbf{B}_0$ term. 

From this, we conclude that if (1) our currents and charges are independent of time and (2) with reasonable assumptions about how rapidly our current distributions vary on length scales $\sim\lambda_a \ll L$, the effect of the axion can be given by $\mathbf{E}_a(\mathbf{x},t)=-\gagg a \mathbf{B}_0$. However, this is \emph{not} the limit proposed for axion searches in the mass range $m_a\lesssim1\,\mu$eV.

It is worth pointing out that the infinite solenoid of the Sec.~\ref{sec:InfiniteSolenoid} does not satisfy the second assumption made here, and is therefore not a special case of this discussion. Rather the infinite solenoid is a particularly germane demonstration of the conclusions reached here in a geometry which can be easily solved with all the important effects reproduced in a single dimension. 
\section{Alternate Approach Using Polarization}
\label{sec:VacuumPolarization}

\begin{figure*}
\centering
\subfloat[\label{fig:polarizedCapacitor}]{\centering\includegraphics[width=.3\textwidth]{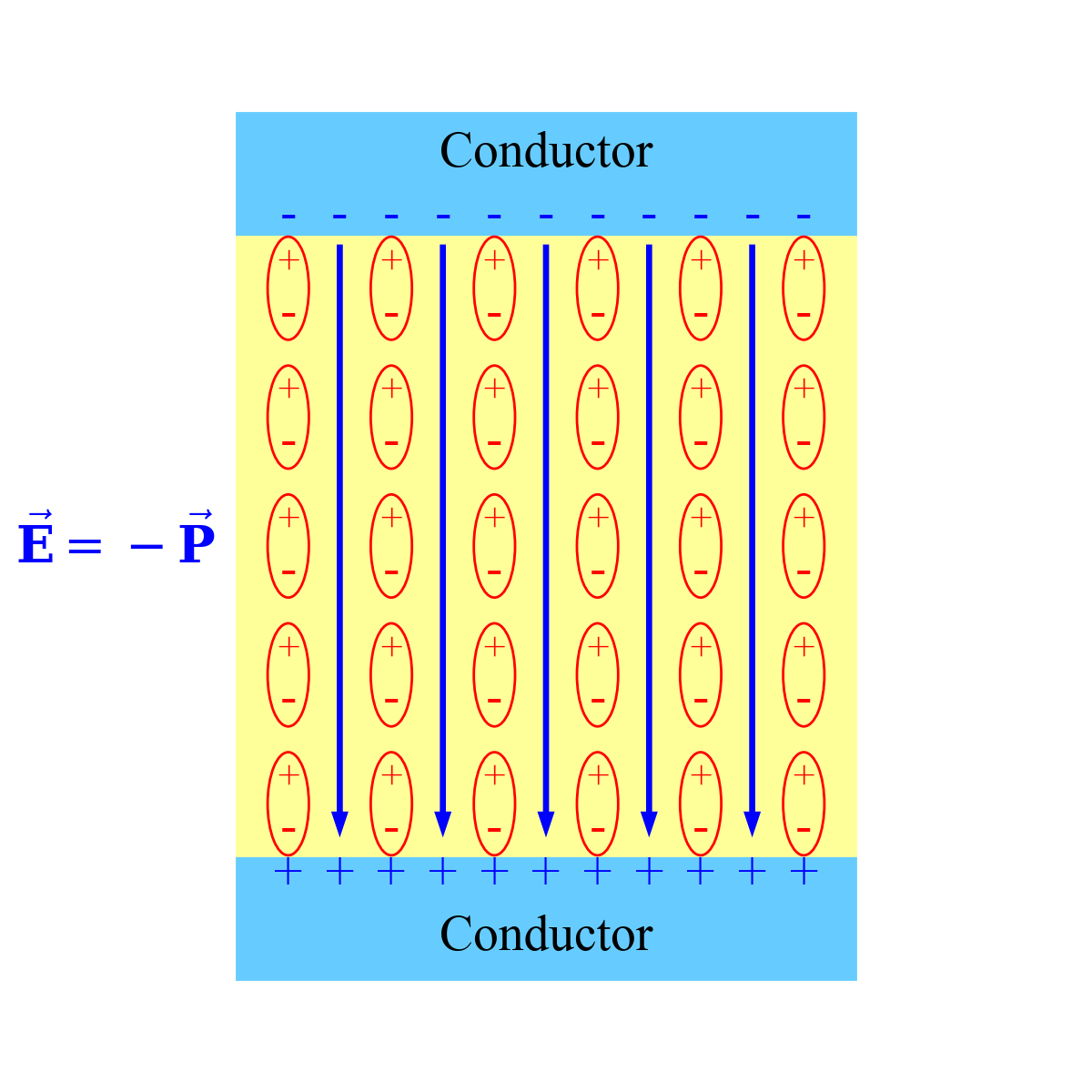}}\hspace{3cm}
\subfloat[\label{fig:polarizedSolenoid}]{\centering\includegraphics[width=.3\textwidth]{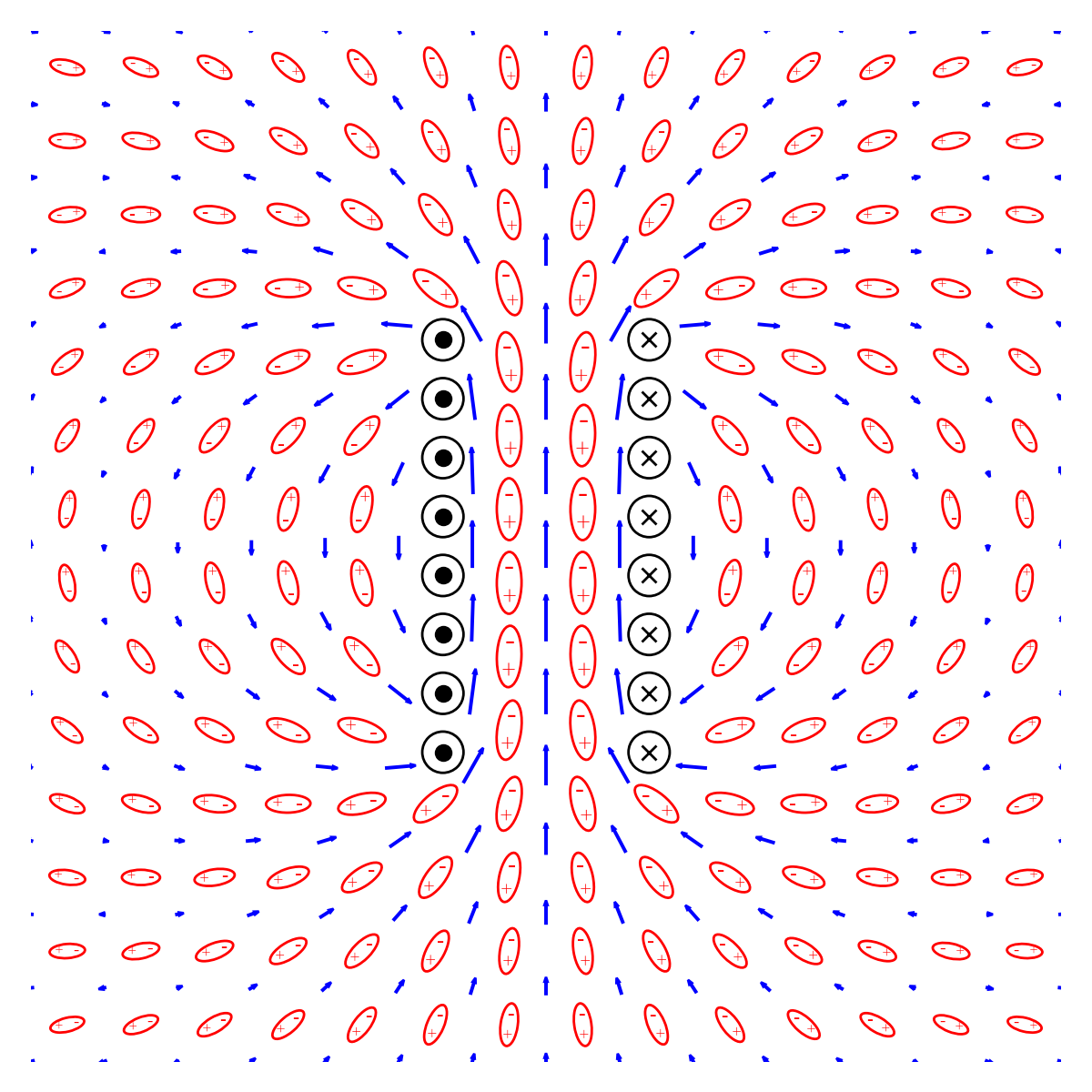}}
\caption{Left: A polarized material like a dielectric (in yellow), placed between two conducting planes. Within the bulk of the material the bound electric dipoles pair off and produce no net field since $\bm{\nabla}\cdot\mathbf{P}=0$. At the surfaces, we have a discontinuity in $\mathbf{P}$ resulting in an overall field and a build up of charge on the conductors. Right: A virtual axion induced polarization $\mathbf{P}_a$ (red) from a magnetic field (blue) from a solenoidal current (black). The divergence of the magnetic field everywhere is zero, so $\bm{\nabla}\cdot\mathbf{P}$ can only be proportional to the gradients of the axion field. Even when a conductor is placed in the field, the $\mathbf{B}$ field and thus $\mathbf{P}$ are divergenceless (up to terms proportional to $\bm{\nabla} a$).}
\label{fig:PolarizationExamples}
\end{figure*}

In the previous sections, we have worked with the vacuum fields $\mathbf{E}$ and $\mathbf{B}$, however, we can extend the entire discussion to the macroscopic formulation using $\mathbf{D}$ and $\mathbf{H}$ fields in the usual way. In this section, we address an approach that incorporates the axion induced effects as a type of vacuum polarization, similar to the polarization of materials. This approach was originally proposed in \cite{BEAST_Paper,Tobar2018}, however those works reach incorrect physical conclusions as the analogy between material and axion induced polarizations is subtle and the correct boundary conditions must be enforced. Nevertheless, this approach is perfectly consistent with the approach in the previous sections.

Following \cite{BEAST_Paper,Tobar2018}, we can reformulate Eqn.~\ref{eqn:modmax} in terms of the macroscopic fields $\mathbf{D}$ and $\mathbf{H}$: 
\begin{subequations}
\begin{eqnarray}
\bm{\nabla}\cdot\mathbf{D}&=&\rho_e+\rho_f+\gagg\mathbf{B}\cdot\bm{\nabla}a\,,\\
\bm{\nabla}\cdot\mathbf{B}&=&0\,,\\
\bm{\nabla}\times\mathbf{E}&=&-\dt{\mathbf{B}}\,,\\
\bm{\nabla}\times\mathbf{H}&=&\mathbf{J}_e+\mathbf{J}_f+\dt{\mathbf{D}}-\gagg\left(\mathbf{E}\times\bm{\nabla}a+\dt{a}\mathbf{B}\right)\,.
\end{eqnarray}
\end{subequations}
These equations are of course identical to Eqn.~\ref{eqn:modmax}, however are more common when including the response of media. Also note that we have explicitly included the response of free charges in the form of $\rho_f$ and $\mathbf{J}_f$, while implicitly including the response of bound charges $\rho_b$ and $\mathbf{J}_b$ in $\mathbf{D}$ and $\mathbf{H}$. In the approach of \cite{BEAST_Paper,Tobar2018} however, we further rewrite this in terms of a set of \emph{modified} fields
\begin{subequations}
\begin{eqnarray}
\mathbf{D}_a &=& \mathbf{D}-\gagg\left(a\mathbf{B}\right)\,,\label{eqn:mod_D_field}\\
\mathbf{H}_a &=& \mathbf{H}+\gagg\left(a\mathbf{E}\right)\,, \label{eqn:mod_H_field}
\end{eqnarray}
\label{eqn:mod_DH_fields}
\end{subequations}
with which we can write an analogous set of macroscopic Maxwell's equations with no axion modification terms
\begin{subequations}
\begin{eqnarray}
\bm{\nabla}\cdot\mathbf{D}_a&=&\rho_f\,,\\
\bm{\nabla}\cdot\mathbf{B}&=&0\,,\\
\bm{\nabla}\times\mathbf{E}&=&-\dt{\mathbf{B}}\,,\label{eqn:modmax_alt_3}\\
\bm{\nabla}\times\mathbf{H}_a&=&\mathbf{J}_f+\dt{\mathbf{D}_a}\,.\label{eqn:modmax_alt_4}
\end{eqnarray}
\label{eqn:modmax_alt}
\end{subequations}

In four-vector notation, what we have done here is to envelope the axion current of Eqn.~\ref{eqn:AxionCurrent} into a redefinition of the electromagnetic field tensor $F^{\mu\nu}\rightarrow F^{\mu\nu}_a=F^{\mu\nu}-P_a^{\mu\nu}$, where 
\begin{equation}
\partial_\mu P_a^{\mu\nu}=J_a^{\nu}.
\end{equation}
We can see that $P_a^{\mu\nu}$ should be given by
\begin{eqnarray}
P_a^{\mu\nu}&=&\gagg a\tilde{F}^{\mu\nu}
\\&=&\gagg a \begin{pmatrix} 0 & -B_x & -B_y & -B_z\\ B_x & 0 & E_z & -E_y \\ B_y & -E_z & 0 & E_x \\ B_z & E_y & -E_x & 0 \end{pmatrix}\,. \nonumber
\end{eqnarray}
And of course, the continuity equation follows trivially from the fact that 
\begin{equation}
\partial_\mu J_a^{\mu} = \gagg\partial_\mu\partial_\nu a \tilde{F}^{\mu\nu} = 0\,,
\end{equation}
because the derivatives are symmetric under interchange of $\mu$ and $\nu$ and $\tilde{F}^{\mu\nu}$ is anti-symmetric.

This entire approach is completely analogous to the way the macroscopic form of Maxwell's equations splits the electric current into $J^\mu_{\rm bound}$ and $J^\mu_{\rm free}$ and attaches the former into a redefinition of $F^{\mu\nu}\rightarrow G^{\mu\nu}=F^{\mu\nu}-P^{\mu\nu}_{\rm bound}$, where
\begin{equation}
P^{\mu\nu}_b=\begin{pmatrix}0& P_x&P_y&P_z\\
-P_x&0&M_z&-M_y\\
-P_y&-M_z&0&M_x\\
-P_z&M_y&-M_x&0\end{pmatrix}\,,
\end{equation}
for a material polarization $\mathbf{P}$ and magnetization $\mathbf{M}$, such that $\partial_\mu P^{\mu\nu}_b=J_b^{\nu}$. In each of these steps, our equations of motion remain completely unchanged and the continuity equation is always satisfied. We are simply moving terms around.
\begin{subequations}
\begin{eqnarray}
\partial_\mu F^{\mu\nu}_a&=&J^\nu_f\,,\\
\partial_\mu G^{\mu\nu}-\partial_\mu P_a^{\mu\nu}&=&J^\nu_f\,,\\
\partial_\mu F^{\mu\nu}-\partial_\mu P_b^{\mu\nu}&=&J_f^{\nu}+J_a^{\nu}\,,\\
\partial_\mu F^{\mu\nu}&=&J^{\nu}_f+J_a^{\nu}+J_b^{\nu}\,.
\end{eqnarray}
\end{subequations}

This appears to be a tidy reformulation of Eqns.~\ref{eqn:modmax}, however, it must be emphasized that \emph{the physics is completely unchanged from the previous sections.} Further, great care has to be taken when using these $\mathbf{D}_a$ and $\mathbf{H}_a$ fields, as the simplicity of Eqns.~\ref{eqn:modmax_alt} can be deceptive. The reason is that the Lorentz force has not been changed, $\mathbf{f}=\rho_e\mathbf{E}+\mathbf{J}_e\times\mathbf{B}$. In other words, charges and currents still rearrange themselves in response to $\mathbf{E}$ and $\mathbf{B}$ fields. Therefore boundary conditions must still be placed on $\mathbf{E}$ and $\mathbf{B}$ rather than on $\mathbf{D}_a$ and $\mathbf{H}_a$.

The purpose of this approach, however, is to continue the analogy, and to write a set of axion polarization and magnetization fields:
\begin{subequations}
\begin{eqnarray}
\mathbf{P}_a &=& -\gagg\left(a\mathbf{B}\right)\,, \label{eqn:beast_polarization} \\
\mathbf{M}_a &=& \gagg\left(a\mathbf{E}\right)\,. \label{eqn:beast_magnetization}
\end{eqnarray}
\label{eqn:beast_fields}
\end{subequations}
But this is where the subtleties become critical. For instance, one must keep in mind that 
\begin{eqnarray}
\bm{\nabla}\cdot\mathbf{P}_a &=& -\gagg\bm{\nabla}\cdot\left(a\mathbf{B}\right)\nonumber\\
&=& -\gagg \left[\bm{\nabla}a\cdot\mathbf{B}+a\bm{\nabla}\cdot\mathbf{B}\right]\nonumber\\
&=&-\gagg\bm{\nabla}a\cdot\mathbf{B}\nonumber\\
&\sim& O(\gagg v_{\rm DM}).
\label{eqn:AxionPolDivergenceless}
\end{eqnarray}
In other words, in the limit of small spatial gradients in $a$, the axion ``bound charge density'' is suppressed. Substituting this into Eqn.~\ref{eqn:modmax1} tells us that $\mathbf{P}_a$ does \emph{not} create an $\mathbf{E}$ field directly. We often intuitively think that an electrically polarized material has an associated electric field. However, this field comes from bound surface charges at the edge of the polarized material. For instance, a dielectric material must be cut to be placed inside of a capacitor, and it is at the boundaries of the dielectric that we have non-zero $\bm{\nabla}\cdot\mathbf{P}$ (see Fig.~\ref{fig:polarizedCapacitor}). But Eqn.~\ref{eqn:AxionPolDivergenceless} indicates that no such a boundary for $\mathbf{P}_a$ exists and so $\bm{\nabla} \cdot \mathbf{P}_a$ is suppressed by $v_{\rm DM}$ (see Fig.~\ref{fig:polarizedSolenoid}). So while it might naively appear that an electric field must be present due to the axion polarization, \emph{it is not}.

Instead a time-varying $\mathbf{P}_a$ generates a time varying magnetic field and that time-varying magnetic field can generate time-varying electric fields. Stepping back to our example of the infinite solenoid, we can easily calculate the polarization and magnetization to first order in \gagg (neglecting terms proportional to $\bm{\nabla} a$):
\begin{subequations}
\begin{eqnarray}
\mathbf{P}_a &=& \begin{cases} -\gagg aB_0\mathbf{\hat{z}}\,, & \rho<R\\ 0 & \rho > R\,, \end{cases}\\
\mathbf{M}_a &=& 0 \qquad\qquad \mathrm{Everywhere}\,.
\end{eqnarray}
\end{subequations}
The intuition would be to view this as a time varying electric field inside our solenoid. But there is no divergence in $\mathbf{P}$ to generate such an electric field. Instead we note that $\mathbf{P}_a$ varies in time with $a$ and plug these values into Eqn.~\ref{eqn:modmax_alt_4} and recover Eqn.~\ref{eqn:modmax4}. This will recover the result in Sec.~\ref{sec:InfiniteSolenoid}.

This underlines the fact that an axion polarization with no space-time derivatives cannot have any physical manifestations. This is also evident in the Lagrangian, as the $aF\tilde{F}$ terms becomes a total derivative in the limit that $\partial_\mu a =0$. An analogous argument can be made about magnetization induced magnetic fields. Despite our intuition otherwise, the magnetization, $\mathbf{M}_a$, alone cannot generate a physically observable magnetic field, only when $\bm{\nabla}\times\mathbf{M}_a\ne0$.

The approach of calculating axion induced polarization and magnetization is completely equivalent to the approach outlined in the first part of this paper. But great care must be taken when using this approach, because subtleties in the application of boundary conditions and physical intuition can conspire to produce physical effects where they should be suppressed.

\section{Conclusion}
In this work, we have stepped through the calculation of the axion induced $\mathbf{E}$ and $\mathbf{B}$ fields in the presence of a strong magnetic field in an infinite solenoid. We showed that the solution $\mathbf{E}=-\gagg a \mathbf{B}$ is part of the full solution of the modified Maxwell's equations, however by itself it does not satisfy the required boundary conditions. Instead the full solution is equivalent to that of a multipole antenna with a current excitation $\mathbf{J}_a=\gagg \partial_t a \mathbf{B}_0$. We then showed that in the large $\lambda_a$ limit, the full solution suppresses vacuum electric fields everywhere by $\left(\frac{R}{\lambda_a}\right)^2$. 

We then laid out the generic derivation of the \MQS approximation in the presence of an axion field and demonstrated that, in any experimental setup with a time-independent charge and current distribution, the axion induced vacuum $\mathbf{E}$ fields are \emph{always} suppressed relative to the axion induced vacuum $\mathbf{B}$ fields in the large $\lambda_a$ limit. The conclusions of this work directly contradict the arguments outlined in \cite{BEAST_Paper,Tobar2018}, and this implies that the limits shown in \cite{BEAST_Paper} are too strong by $\sim$6.5 orders of magnitude. However, it is equally important to point out that the $\bm{\nabla} a$ effects, which were ignored in this work, are only suppressed by three orders of magnitude. The approach proposed in \cite{BEAST_Paper,Tobar2018} may be a powerful ``wind'' experiment searching for axion induced effects through the $\bm{\nabla} a$ terms.

Finally, it should be noted that these conclusions are based on the vacuum solutions of the $\mathbf{E}$ and $\mathbf{B}$ fields. These fields can be further shaped through the placement of conductors and free charges, which can mix these fields. For example, placing an inductor in a time varying $\mathbf{B}$ field will produce an $\mathbf{E}$ field in the inductor in the usual way, which will not be suppressed by additional powers of $\lambda_a$. This also underscores the need for all axion haloscopes to carefully analyze the effect of the boundaries of their fields.

\emph{Note:} A recent paper \cite{Beutter:2018xfx} has redone the calculation from \cite{HONG1991} without assuming a homogeneous $\mathbf{B}$ field. The results in that paper agree with the results presented here, but are achieved with an elegant field theory approach. Another paper \cite{Kim:2018sci} has performed a similar Taylor expansion of the fields to calculate solutions inside conducting cavities.  
\begin{acknowledgements}

The authors would like to thank Yoni Kahn, Reyco Henning, Lindley Winslow, Jesse Thaler, and Hongwan Liu for the huge amount of useful input and suggestions in writing this paper. We would like to further acknowledge the fruitful discussions with Kent Irwin and Aaron Chou at the 3$^{\rm rd}$ Workshop on Microwave Cavities for Axion Detection that were the starting point for this work. Finally, we would like the acknowledge the friendly discussions with Michael Tobar which helped clarify the differences between our approaches. This work was supported by the NSF under award number 1806440. \end{acknowledgements}

\end{document}